\documentclass[a4paper]{article}  
\usepackage{indentfirst}  
\usepackage{bm}    
\usepackage{graphicx}  
\usepackage{url}
\usepackage{fancyhdr}

\usepackage{geometry}
\usepackage{amstext}
\usepackage{amsmath}
\begin{document}    
\thispagestyle{empty} \vspace*{0.8cm}\hbox
to\textwidth{\vbox{\hfill\noindent \\ \textit{Proceedings of the 8th International Conference on Pedestrian and Evacuation Dynamics (PED2016)\\
Hefei, China - Oct 17 -- 21, 2016\\
Paper No. 25}
\hfill}}
\par\noindent\rule[3mm]{\textwidth}{0.2pt}\hspace*{-\textwidth}\noindent
\rule[2.5mm]{\textwidth}{0.2pt}

\begin{center}
\LARGE\bf The effect of social roles on group behaviour
\end{center}

\begin{center}
\rm Francesco Zanlungo$^{1,2}$, Zeynep Y\"ucel$^{2,3}$ and Takayuki Kanda$^{2}$
\end{center}

\begin{center}
\begin{small} \sl
${}^{\rm 1}$ Kingston University, London, UK \\  
${}^{\rm 2}$ ATR International, Kyoto, Japan \\  
${}^{\rm 3}$ Research fellow, JSPS, Tokyo, Japan \\
zanlungo@atr.jp, zeynep@atr.jp, kanda@atr.jp
\end{small}
\end{center}
\vspace*{2mm}

\begin{center}
\begin{minipage}{15.5cm}
\parindent 20pt\small
\noindent\textbf{Abstract - } In a recent series of papers, we proposed a mathematical model for the dynamics of a group of interacting pedestrians.
The model is based on a non-Newtonian potential, that accounts for the need of pedestrians to keep both their interacting partner and their walking goal in their vision field, and to keep a comfortable distance between them.
These two behaviours account respectively for the angular and radial part of the potential from which the force providing the pedestrian acceleration is derived. The angular term is asymmetric, i.e. does not follow
the third law of dynamics, with observable effects of group formation and velocity.
We first assumed the group to move in a scarcely dense environment, whose effect could be modelled through a ``white noise thermic bath'', and successfully compared the predictions of the model with 
observations of real world pedestrian behaviour. We then studied, both from an empirical and a theoretical standpoint, the effect of crowd density on group dynamics. We verified that the average effect of crowd density may be 
modelled by adding a harmonic term to the group potential. The model predictions, which include ``phase transitions'' in the group configuration (e.g. in 3 people groups transition from a ``V'' formation to a ``$\Lambda$'' one,
and eventually to pedestrians walking in a line), are again confirmed, at least in the observed density range, by a comparison with real world data.
Until now we had averaged all pedestrian data collected in a given environmental setting (i.e. in corridors of similar width and at similar crowd densities) without differentiating on group composition and social roles.
In this work, we present preliminary results on these features, namely we study how the group configuration and velocity is affected by inter-pedestrian relation (family, couples,
colleagues, friends), purpose (work, leisure) and gender.
We also show results related to the effect of asymmetric interactions, that confirm further the non-Newtonian nature of gaze-based angular interaction in our model.
\end{minipage}
\end{center}

\begin{center}
\begin{minipage}{15.5cm}
\begin{minipage}[t]{2.3cm}{\bf Keywords:}\end{minipage}
\begin{minipage}[t]{13.1cm}
Group behaviour, social interaction, experimental data, mathematical modelling 
\end{minipage}\par\vglue8pt
\end{minipage}
\end{center}

\section{Introduction}
Urban crowds are characterised by the presence of a large number of social groups. The ratio between individual pedestrians and pedestrians moving in groups
may change considerably between different environments and at different times of the day \cite{M,M2,OOC}, but it is in general never negligible, with groups representing up to 85\% of the walking population \cite{schultz2,mou2}.
Despite this empirical evidence about the importance of groups, the standard approach in microscopic (agent-based) pedestrian modelling is to assume that the crowd is composed of individuals,
moving without any preferential ties to other pedestrians. Such an approach is, at least, a strong simplification as the description of a system of complex molecules as a monoatomic gas and,
although justified in the infancy of the field, it should be now replaced by more realistic models taking in account group and in general social interactions.

Indeed, in recent years, a few studies concerning empirical observations and mathematical modelling of the groups' characteristic configuration and velocity \cite{M,schultz2,mou2,costa,zan3,dyads,koster,kara,zhang}
have been introduced. In a recent series of papers \cite{M,M2,M3}, we focused on the development of a mathematical model to describe group interaction. The model is based on few and intuitive ideas 
about social interaction in pedestrian groups, and its predictions are in agreement with the observed natural behaviour of pedestrians. It may also describe
how group behaviour is modified by crowd density, just by adding an harmonic term to the low density potential.

In this work, after reviewing our model, we present some new empirical findings regarding how group dynamics is affected by group composition (purpose, inter-group relations, gender) and by asymmetric interactions,
and discuss these findings under the light of our mathematical framework.
\section{Potential for group interaction}
\begin{figure}[tb!]
\begin{center}
\includegraphics[width=0.25\linewidth]{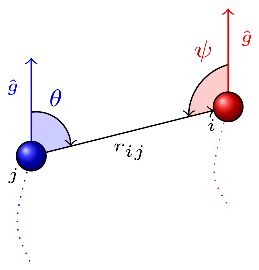} \hspace{0.1\linewidth}\includegraphics[width=0.6\linewidth]{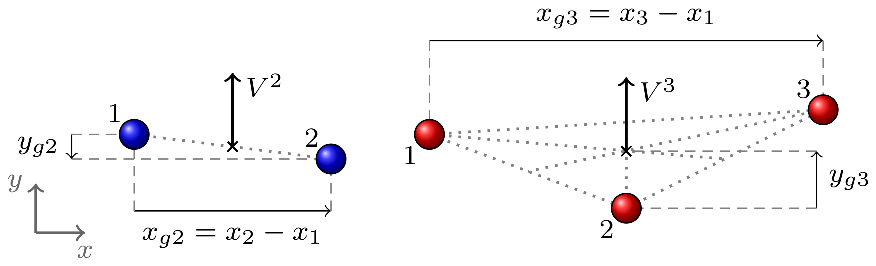}
\caption{Left: $r$ and $\theta$ give the position of pedestrian $i$ with respect to $j$. $|\theta|$
is the angle that $j$'s gaze has to span between the goal direction $\hat{\mathbf{g}}$ 
and $i$, while the corresponding angle for $i$ is given by $|\psi|$. Right: Definition of the 2 and 3 people group observables. }
\label{f1}
\end{center}
\end{figure}
\subsection{Mathematical formulation}
Following \cite{M}, we assume that 2 pedestrians, identified as $i$ and $j$, are socially interacting while walking towards a common goal, identified by a unit vector $\hat{\mathbf{g}}$. 
Their relative position, $\mathbf{r}\equiv\mathbf{r}_{ij}\equiv \mathbf{r}_i-\mathbf{r}_j$, 
may be written in polar coordinates as $(r,\theta)$, where $-\pi < \theta \leq \pi$ is the angle between $\mathbf{r}_{ij}$ and $\hat{\mathbf{g}}$ (see Fig. \ref{f1}).  We also define the angle $\psi$
\begin{equation}
\psi \equiv \left\{
\begin{array}{rl}
\theta-\pi & \text{if } \theta >  0,\\
\theta + \pi & \text{if } \theta \leq 0,
\end{array} \right.
\end{equation}
so that $|\psi|$ is the magnitude of the angle that $i$'s gaze has to span between the goal and the interaction partner, while $|\theta|$ is the corresponding angle for $j$.
Our model is based on the assumption that pedestrians want to look towards their goal for safe walking, but for social interaction they want both to have their partner in their vision field \cite{knapp},
and to be visible to their partner, so that their gazes may meet \cite{kleinke}. When pedestrians are in a relative position that does not allow for an optimal interaction, they feel
a {\it discomfort}. We model this discomfort and assume that it plays the same role that potential energy plays in physics, namely the pedestrian's acceleration is determined by 
the negative gradient of the potential, so that the pedestrians move towards the {\it discomfort minimum}.
The discomfort should grow with the angles  $|\theta|$ and $|\psi|$, and we may assume, out of simplicity, the law to be quadratic
\begin{equation}
\label{angular}
\Theta^{\eta}(\theta) = (1+\eta) \theta^2+(1-\eta) \psi^2,\;\;\;\;\;\;\;-1 \leq \eta \leq 1.
\end{equation}
The discomfort potential should have also a radial component, assuming a minimum at the distance $r_0$ at which social interaction is maximally comfortable \cite{Z}. A simple choice is to have
this potential to diverge for $r \to 0$ to avoid physical overlapping, and to grow linearly for $r \to \infty$, so that the interaction force converges to a constant value for large distances. 
A simple choice\footnote{Describing well pedestrian behaviour close to the potential minimum, but failing far from the minima, due probably also to the fact that far away pedestrians stop interacting.} is
\begin{equation}
\label{radiale}
R(r)= \frac{r}{r_0}+\frac{r_0}{r}.
\end{equation}
We assume the total potential to be the sum of the angular and radial term\footnote{So that the statistical distributions of $r$ and $\theta$ will be independent, as the velocity and position in a classical statistical gas.}
\begin{equation}
\label{potenziale}
U^\eta_{ij}(r,\theta)= C_r R(r) + C_\theta \Theta^\eta(\theta).
\end{equation} 
The acceleration of pedestrian $i$ results to be
\begin{equation}
\label{grad}
\dot{\mathbf{v}}_i=\mathbf{F}_{ij}=-\mathbf{\nabla}_i U^\eta_{ij}(\mathbf{r}_i-\mathbf{r}_j).
\end{equation}
If we choose a Cartesian system with $y$ axis aligned to the goal direction $\hat{\mathbf{g}}$ we have
\begin{equation}
\label{f}
F^\eta_x= \frac{C_r}{r_0} \left(\frac{r^2_0}{r^2}-1\right) \sin \theta - \frac{4}{r} C_\theta (\theta-\theta_{\text{sgn}({\theta})}) \cos \theta,\;\;\;\;\;\;\;\;F^\eta_y= \frac{C_r}{r_0} \left(\frac{r^2_0}{r^2}-1\right) \cos \theta + \frac{4}{r} C_\theta (\theta-\theta_{\text{sgn}({\theta})}) \sin \theta,
\end{equation} 
where $\theta_{\pm}$ are the angles at which the angular potential attains a minimum,
\begin{equation}
\label{tpm}
\theta_{\pm}=\pm (1-\eta) \frac{\pi}{2}.
\end{equation}
This system has some properties that make its dynamics very different from the one of a classical Newtonian $n$ body system. Indeed,
 $U^\eta_{ij}(\mathbf{r}_i-\mathbf{r}_j)$ is not the potential of the $(i,j)$ system, but only of the pedestrian $i$. For
$\eta \neq 0$ we have 
\begin{equation}
U^\eta(-\mathbf{r})\neq U^\eta(\mathbf{r})\qquad \Rightarrow \mathbf{F}_{ij} \neq -\mathbf{F}_{ji}
\end{equation}
in contradiction with Newton's third law and conservation of momentum, as may be expected for an interaction based on gaze \cite{Turchetti}. It is possible to show that the centre of mass of the  $(i,j)$ 
system\footnote{Here we assume ${0 < \theta \leq \pi}$ to be the angle giving the position of the pedestrian on the right.} is accelerated by purely internal forces, with
\begin{equation}
\label{derV}
\dot{V}^\eta_x= -\eta \frac{2 \pi}{r} C_\theta \cos \theta,\qquad\dot{V}^\eta_y= \eta \frac{2 \pi}{r} C_\theta \sin \theta.
\end{equation}
On the other end, the relative dynamics is given by a Newtonian potential
\begin{equation}
\label{relative}
\dot{\mathbf{v}}\equiv \ddot{\mathbf{r}}_i-\ddot{\mathbf{r}}_j=- 2 \mathbf{\nabla} U^0(\mathbf{r}).
\end{equation}
We expect this model to describe correctly pedestrian behaviour close to  potential minima. In a wide, low density environment
pedestrians should be close to these minima, and we may model the interaction with the environment using white noise $\mathbf{\Xi}$
with standard deviation $\sigma$. Using the Social Force Model framework \cite{Hel},
we assume pedestrians to be dragged towards the goal with acceleration\footnote{We use the value $\kappa=1.52$ s$^{-1}$ from \cite{zan4}. $v^{(1)}$ is the preferred velocity of pedestrians when walking alone.}
\begin{equation}
\label{drag}
\mathbf{F}^g_i=\kappa (\mathbf{v}_p-\mathbf{v}_i),\qquad \mathbf{v}_p=v^{(1)} \hat{\mathbf{g}}.
\end{equation}
As a result the dynamics of the centre of mass for a 2 people group is, ignoring stochastic terms,
\begin{equation}
\label{dycm}
\dot{\mathbf{V}}=\kappa (\mathbf{v}_p-\mathbf{V})+\dot{\mathbf{V}}^\eta,
\end{equation}
while the relative dynamics is given by the stochastic equation
\begin{equation}
\label{dineqcm}
\dot{\mathbf{v}}=-\kappa \mathbf{v}- 2 \mathbf{\nabla} U^0(\mathbf{r})+\mathbf{\Xi}.
\end{equation} 
The probability distribution function for the relative distance between pedestrians in a group of 2 is given by the Boltzmann distribution
\begin{equation}
\label{boltz}
\rho(\mathbf{r})\propto \text{exp}(-\beta U^0(\mathbf{r})).
\end{equation}
\subsection{Consequences of the model}
In the model above we assumed the interaction between the two pedestrians to be determined by potentials with the same functional form, a point that we will better study in this work. Under this assumption, 
the equilibrium configuration is necessarily an abreast one, as determined by eq. (\ref{relative}), in which $\eta$ does not appear. The effect of the non-Newtonian
interaction is anyway observable, since the two pedestrians are not in their most comfortable configuration and their velocity is slowed down (assuming $\eta<0$, as found by comparison
to the data) by the  {\it load of interaction}
\begin{equation}
\label{deltav2}
\Delta v^{(2)}\equiv v^{(2)}-v^{(1)}=\eta\, C_\theta \,\frac{2 \pi}{r \kappa}  \equiv \frac{\Delta\, a^{(2)}}{\kappa}, 
\end{equation}
$v^{(2)}$ being the velocity of a group of two interacting pedestrians.

In \cite{M} we assumed the potential to act only between first neighbours. Assuming this to be true, if 3 pedestrians were walking in an abreast formation,
the central one would be slowed down by a factor $2 \Delta a^{(2)}$, compared
to the factor $\Delta a^{(2)}$ of the pedestrians on the wings. As a result, the pedestrians walk in a ``V'' formation (the central one being on the rear), with velocity
\begin{equation}
\label{deltav3}
\Delta v^{(3)}\equiv v^{(3)}-v^{(1)}\approx\frac{4}{3}\,\eta \,C_\theta \,\frac{2 \pi}{r \kappa} =\frac{4}{3}\,\Delta v^{(2)}.
\end{equation}  
\subsection{Calibration and validation}
\begin{figure}[tb!]
\begin{center}
\includegraphics[width=0.4\linewidth]{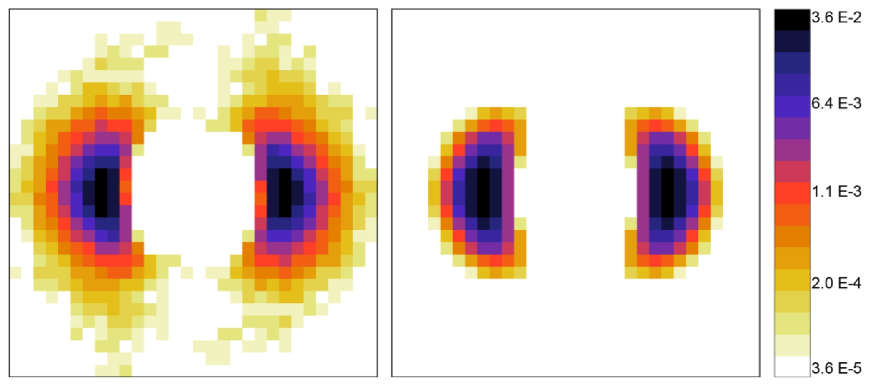}\hspace{0.1\linewidth}\includegraphics[width=0.4\linewidth]{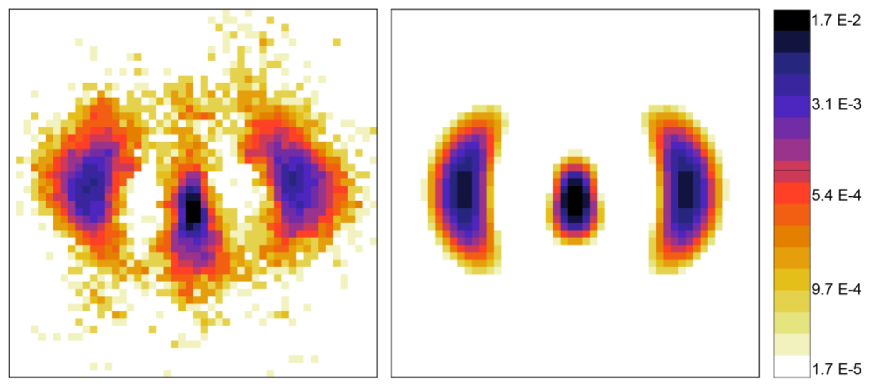}
\caption{Left: comparison, in logarithmic scale, between the empirical (left) and modelled (right) probability distribution function for the position of pedestrians in a 2 people group (positions are given in the centre of mass frame, 
the goal is given by the $y$ axis). Right: comparison, in logarithmic scale, between the empirical (left) and modelled (right) probability distribution function for the position of pedestrians in a 3 people group (same
referential frame as left.). }
\label{f3}
\end{center}
\end{figure}

We compared these predictions with the behaviour of actual pedestrians, 
tracked using 2D laser range sensors \cite{dylan} in a pedestrian
facility in Osaka, and whose group relations were identified by two human coders. The environment did fit well to our hypotheses by being relatively large (a corridor of $\approx 6-7$ meters width) and at low density,  
 ($\approx 0.03$ ped/m$^2$). The model agreed in a qualitative and quantitative way with observations, both regarding the group structure and the group velocity. In particular, given the
individual pedestrian velocity $v^{(1)}=1.336 \pm 0.002$ m/s, and the two people group velocity $v^{(2)}=1.159 \pm 0.006$ m/s, the model predicts $v^{(3)}=1.098$ m/s, in agreement with the observed value
$v^{(3)}=1.110 \pm 0.016$. Fig. \ref{f3} shows, in logarithmic scale, a comparison between the empirical and modelled probability distribution functions for 2 pedestrian groups (left),
and 3 pedestrian groups (right).
\subsection{New evidence for non-Newtonian gazing interaction}
\begin{figure}[tb!]
\begin{center}
\includegraphics[width=0.33\linewidth]{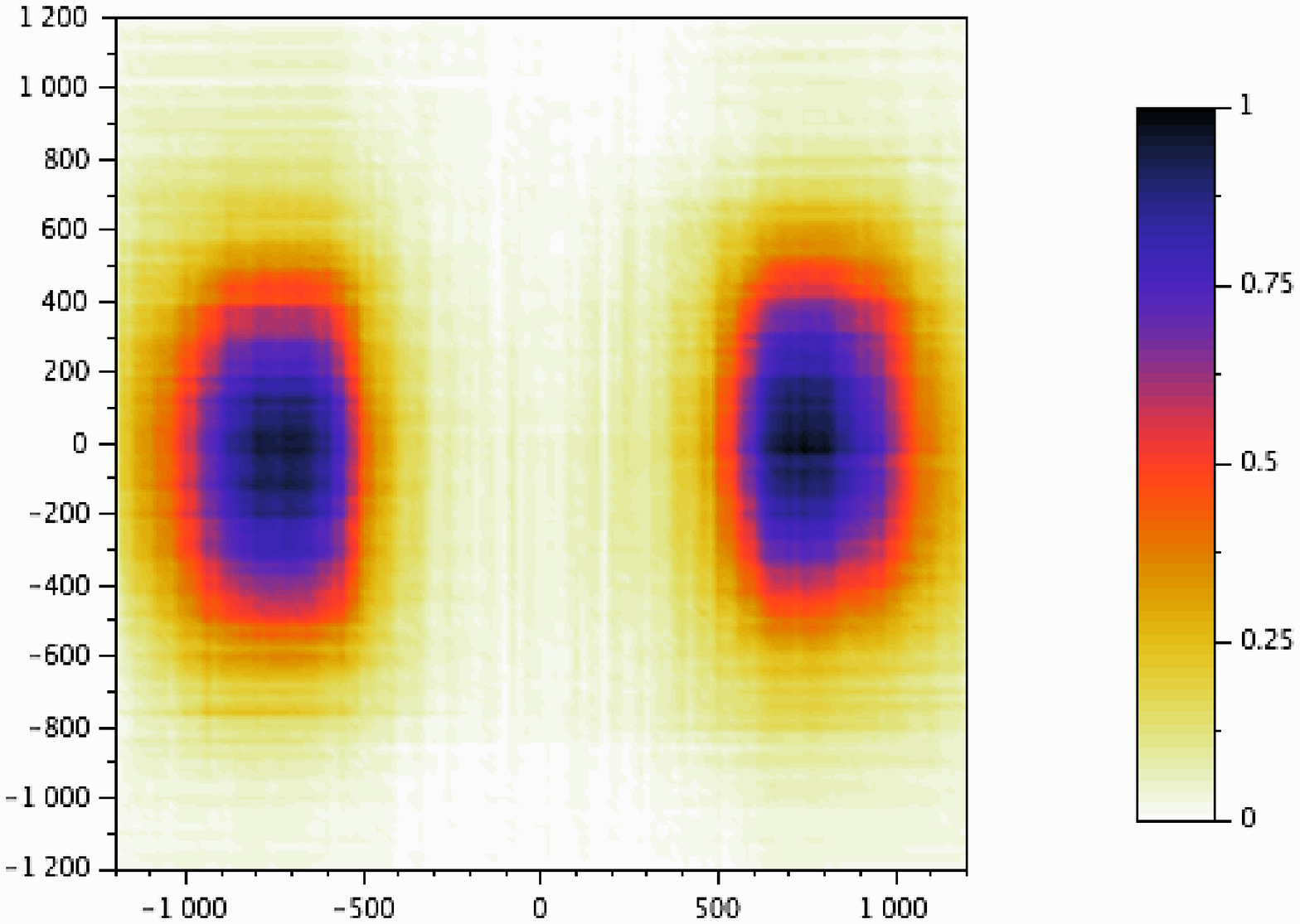}\includegraphics[width=0.32\linewidth]{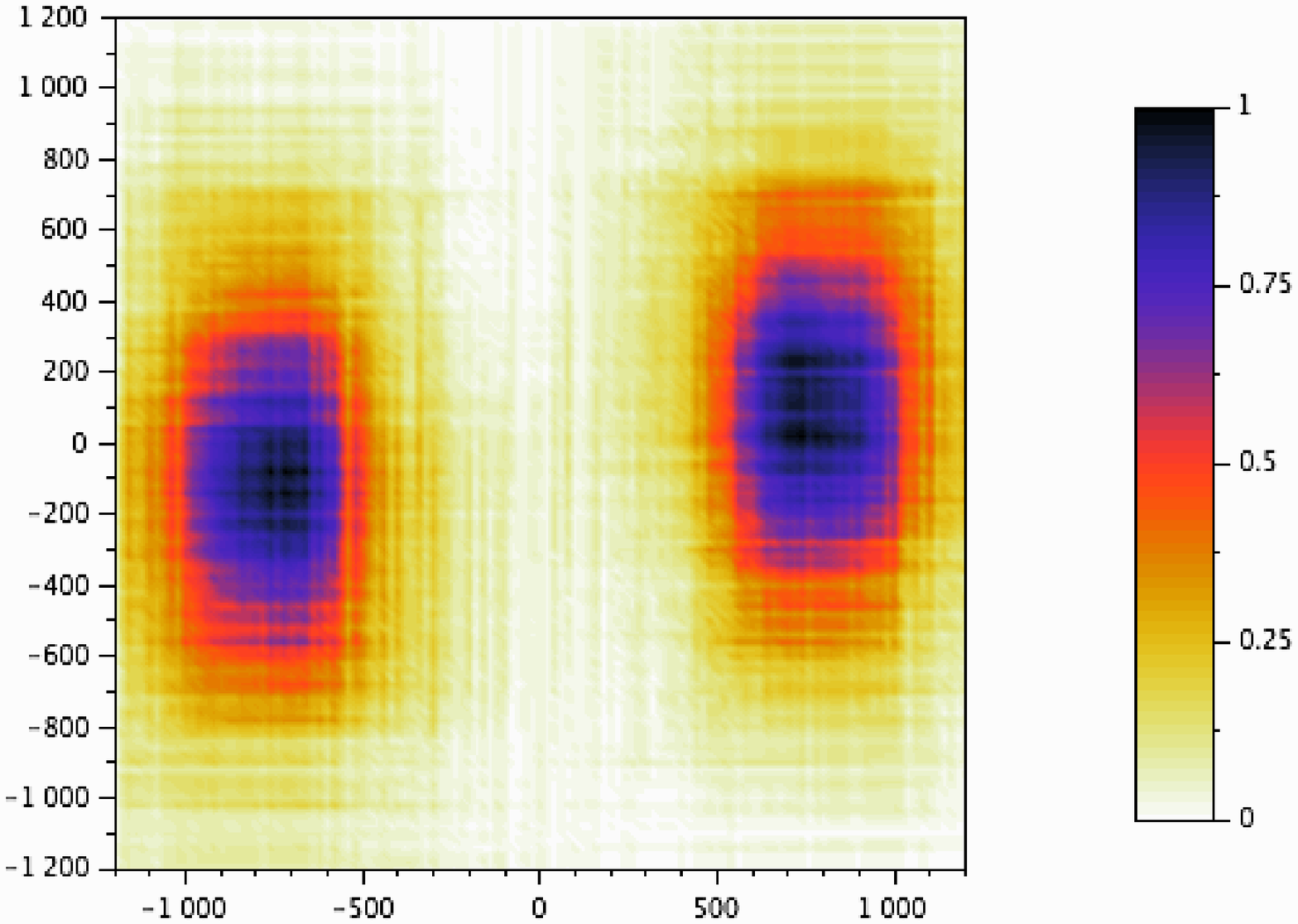}\includegraphics[width=0.33\linewidth]{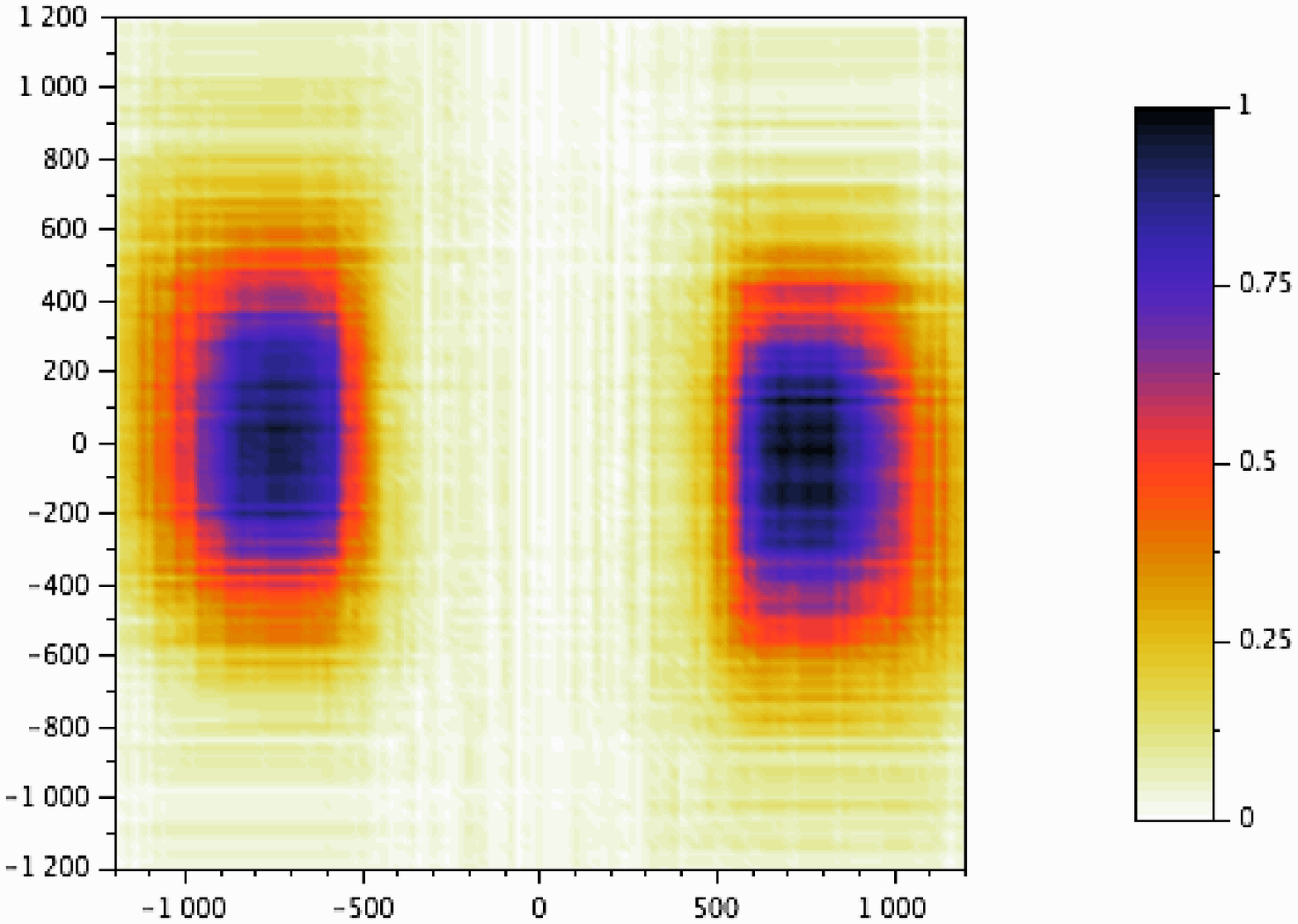}
\caption{Comparison between the empirical probability distribution functions for the position in a 2 people group for pedestrians that are: gazing at each other (left); only the pedestrian on the left gazes at the one on
the right (centre); only the pedestrian on the right gazes at the one on the left (right).}
\label{f5}
\end{center}
\end{figure}

If the slower velocity of groups is due to the $\eta$ term in the potential, in cases in which only a pedestrian is looking at the other, we should see an asymmetric situation,
with the gazing pedestrian walking on the rear. We asked coders to analyse cases in which only one of the pedestrians was watching the other, and our preliminary analysis confirms the prediction (Fig. \ref{f5}).
\section{The effect of density}
In \cite{M2} we presented a large amount of data regarding the change in group shape and velocity with crowd density, and in  \cite{M3} we proposed a mathematical model that may reproduce and explain this behaviour.
\subsection{Definition of observables}
To analyse how group behaviour changes with density we studied, along with the group velocities $v^{(2)}$ and $v^{(3)}$, the group extension along and orthogonal to the direction of motion.

We first define, for a group of $n_g$ pedestrians, the position relative to the group centre
\begin{equation}
\mathbf{r}_i \equiv \mathbf{x}_i-\mathbf{X},\qquad \mathbf{X} \equiv \left(\sum_{i=1}^{n_g} \mathbf{x}_i\right)/n_g.
\end{equation}
We assume that the goal of the group is identified by the group velocity
\begin{equation}
\label{vgroup}
\hat{\mathbf{g}}=\frac{\mathbf{V}}{V}, \qquad \mathbf{V} \equiv \left(\sum_{i=1}^{n_g} \mathbf{v}_i\right)/n_g \qquad V=|\mathbf{V}|
\end{equation}
and compute for each pedestrian the clockwise angle $\theta_i$ between $\hat{\mathbf{g}}$ and $\mathbf{r}_i$. Finally, we define the projections of each $\mathbf{r}_i$
along the goal and orthogonal to it
\begin{equation}
y_i \equiv r_i \cos \theta_i,\qquad x_i \equiv r_i \sin \theta_i.
\end{equation}
Pedestrians are then re-numbered such as 
\begin{equation}
x_1 \leq x_2 \leq \hdots \leq x_{n_g}
\end{equation}
The observables are those features that we verified to change with density in \cite{M2}, namely, along with the group velocity $V=v^{(2)}, v^{(3)}$, 
the 2 and 3 people abreast extension, and the 2 and 3 people extension in the direction of motion (see also Fig. \ref{f1} on left)
\begin{equation}
\label{obs}
x_{g2} \equiv x_2-x_1, \quad y_{g2} \equiv y_2-y_1, \quad x_{g3} \equiv x_3-x_1, \quad y_{g3} \equiv (y_3+y_1)/2-y_2.
\end{equation}
\subsection{The model}
We assumed that the effect of crowd density is a linear recall force with components
\begin{equation}
F^x_i=- K^x(\rho) \frac{x_i}{r_0^2}, \qquad F^y_i=- K^y(\rho) \frac{y_i}{r_0^2}, \qquad K^{x,y}(\rho)\geq 0
\end{equation}
or, equivalently, given by the potential
\begin{equation}
\label{potplus}
U_{\text{macro}}(r,\theta) \equiv \frac{1}{4} \left( K^x(\rho) \frac{r^2 \sin^2 \theta}{r_0^2} + K^y(\rho) \frac{r^2 \cos^2 \theta}{r_0^2} \right).
\end{equation}
\subsection{Calibration and second order interaction}
The data calibration and evaluation was based on the data set \cite{data}, obtained through an automatic tracking system \cite{Drz} and analysed by a human coder to provide the ground truth for social interactions.
The calibration has been performed by choosing the parameters that better reproduced the $\rho$ dependence of $v^{(2)}$ and $x_{g2}$, which resulted to be
\begin{equation}
K^y(\rho) \approx 0,\qquad K^x(\rho) \approx \beta \rho, \quad \beta=2.7 \text{m}^4 \text{ ped}^{-1} \text{ s}^{-2}.
\end{equation}
Namely, the effect of crowd density may be modelled as a harmonic term only in the $x$ direction.

We then compared the prediction of the model regarding the 3 people observables with the data. The agreement was very good concerning $v^{(3)}$ and $y_{g3}$, but poor with respect to $x_{g3}$.

\begin{figure}[h!]
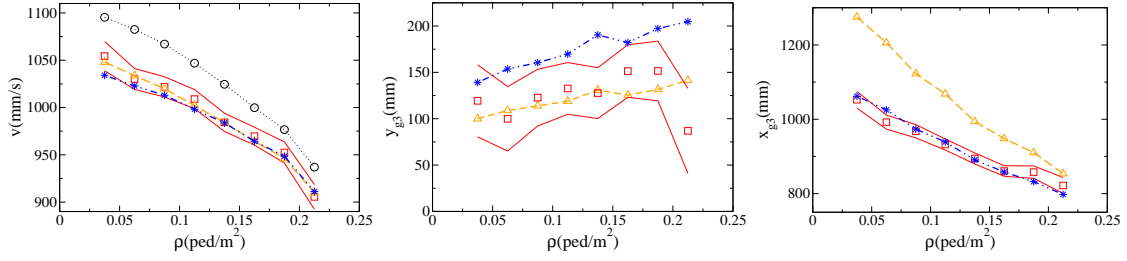

\begin{center}   
\includegraphics[width=0.3\linewidth]{f6.eps} \includegraphics[width=0.3\linewidth]{f7.eps} \includegraphics[width=0.3\linewidth]{f8.eps}
\caption{Left: $\rho$ dependence of $v^{(3)}$ (red squares, confidence intervals in continuous line: empirical average;
dashed orange and triangles: model prediction in absence of second order interactions; dashed-dotted blue and stars: model prediction using the second order interaction of 
eq.~\ref{second}; dotted black and circles shows the empirical $v^{(2)}(\rho)$). Centre: $\rho$ dependence of $y_{g3}$ (red squares: empirical average;
dashed orange and triangles: model prediction in absence of second order interactions; dashed-dotted blue and stars: model prediction using the second order interaction of 
eq.~\ref{second}). Right: $\rho$ dependence of $x_{g3}$ (red squares: empirical average;
dashed orange and triangles: model prediction in absence of second order interactions; dashed-dotted blue and stars: model prediction using the second order interaction of 
eq.~\ref{second}).
}
\label{f6}
\end{center}
\end{figure}

This poor reproduction of the 3 people group width was a problem that we had already identified in \cite{M}, and suggested that it may due to our choice of ignoring any interaction between second neighbours.
We verified that by introducing a force between pedestrians on the wings
\begin{equation}
\label{second}
\mathbf{F}^{\text{second}}= \alpha \frac{C_r}{r_0} \left(\frac{r^2_0}{r^2}-1\right) \frac{\mathbf{r}}{r},
\end{equation} 
where  $\mathbf{r}$ is the distance between pedestrians 1 and 3, and  $\alpha$ was numerically calibrated to $\alpha=0.558$, the agreement between empirical data and model could be largely improved (Fig. \ref{f6}).
\subsection{Phase transitions at high densities}
\begin{figure}[h!]
\begin{center}   
\includegraphics[width=0.4\linewidth]{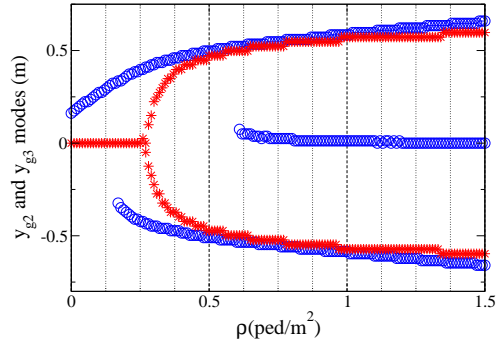}
\caption{Red stars: position of the $y_{g2}$ distribution maxima as a function of $\rho$. 
Blue circles: position of the $y_{g3}$ distribution maxima as a function of $\rho$.} 
\label{f9}
\end{center}
\end{figure}

One of the interesting features of the model is its prediction of density dependent ``phase transitions'' in the group configuration. These phase transitions may be identified as the change (``symmetry breaking'') in the number of
maxima assumed by the $y_{g2}$ distribution (one when pedestrians walk abreast, two when the walk in a line) and by the $y_{g2}$ distribution (one maximum when only the ``V'' configuration is present,
two maxima when also the ``$\Lambda$'' configuration is present, and three maxima when pedestrians walk in a line). The number of maxima predicted by the model is shown in Fig. \ref{f9}. The V-$\Lambda$ transition has been 
qualitatively observed in the data.
\section{The effect of group composition}
\subsection{Method}
We asked 3 different ``coders'', i.e. people not aware of the purpose of our study, to examine video-recordings corresponding to a few hours of our pedestrian group data sets, and to mark pedestrians in
groups according to their apparent gender, age and purpose for visiting the area (work or leisure). They were also asked to specify their apparent social relation, such as family members, couple\footnote{The Japanese word used,
{\it koibito}, usually designs an unmarried, young couple.}, friends and colleagues. All the coders were Japanese nationals, and, although their coding does not provide any objective knowledge regarding group composition, it should 
represent a reliable source of information, which may be complemented with the height of pedestrians, measured by our tracking system \cite{Drz}.
In this preliminary study we used only the observation of a single coder, an experienced non-technical member of our staff that already contributed to the group coding used in \cite{M}. Furthermore, we used only information regarding
the relation, purpose and gender composition\footnote{We studied all female vs all males groups. Mixed groups behaviour is extremely different between, for example, couples and colleagues, and its analysis is left for future work.} of 
two people groups. The comparison between the different coders, the analysis of larger groups, and of the effect of continuous variables such as height and age is left for an upcoming journal paper.
\subsubsection{Observables}
We used as observables the group velocity $V=v^{(2)}$ (eq. \ref{vgroup}), the distance between pedestrians $r$ and the abreast distance $x=x_{g2}$ (eq. \ref{obs}).

For each observable $o$ we compute the average $\mu_i$ for each group $i$, and then provide
its average value as
\begin{equation}
<o> \pm \varepsilon,\qquad <o>=\frac{\sum_{i=1}^N \mu_i}{N}, \qquad \varepsilon=\frac{\sigma}{\sqrt{N}},\qquad \sigma=\sqrt{\left(\sum_{i=1}^N \mu_i^2\right)/{N}-<o>^2}
\end{equation}
where $N$ is the number of groups corresponding to each category (females or males, etc.).
\subsection{Results}
\subsubsection{The effect of gender}
\begin{figure}[tb!]
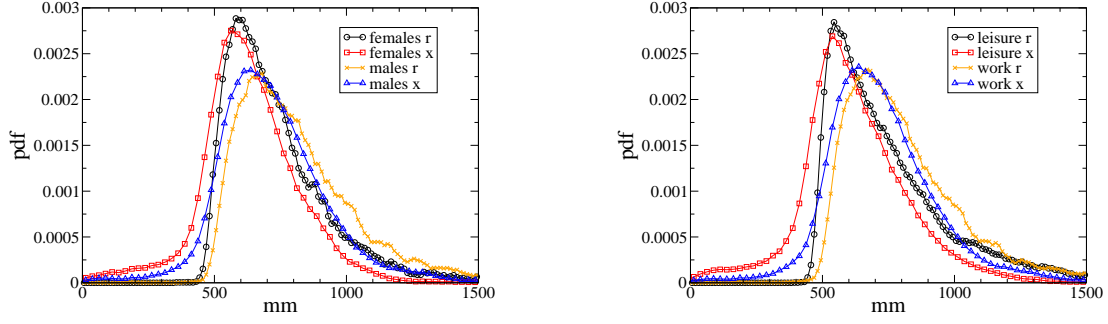

\begin{center}
\includegraphics[width=0.4\linewidth]{f10.eps}\hspace{0.1\linewidth}\includegraphics[width=0.4\linewidth]{f11.eps}
\caption{Left: probability distribution function for the $r$ and $x$ variables in the all females and all males groups.
Right: Probability distribution function for the $r$ and $x$ variables in the groups whose purpose was leisure or work.}
\label{f10}
\end{center}
\end{figure}

Table \ref{T1} compares the values of all observables between groups composed by two males and groups composed by two females, while Fig. \ref{f10} on the left shows the probability distributions of the corresponding spatial variables.
\begin{table}[!ht]
\small
\caption{Observable dependence on gender for 2 people groups.}
\label{T1}
\begin{center}
\begin{tabular} {|c|c|c|c|c|}
	\hline
       &  $N$ &  $V$ &    $r$ & $x$  \\
	\hline
  Females &  57 & $1065 \pm 31$ mm/s  &  $766 \pm 30$ mm & $632 \pm 14$ mm\\
	\hline
  Males  &  128 & $1259 \pm 16$ mm/s  &  $836 \pm 17$ mm & $728 \pm 13$ mm\\
	\hline
\end{tabular}
\end{center}
\end{table}
\subsubsection{The effect of purpose}
Table \ref{T2} compares the values of all observables between groups whose apparent purpose was leisure and groups whose apparent purpose was work, while Fig. \ref{f10} on the right shows the probability distributions of the corresponding 
spatial variables.
\begin{table}[!ht]
\small
\caption{Observable dependence on purpose for 2 people groups.}
\label{T2}
\begin{center}
\begin{tabular} {|c|c|c|c|c|}
	\hline
       &  $N$ &  $V$ &    $r$ & $x$  \\
	\hline
  Leisure &  138 & $1071 \pm 18$ mm/s  &  $806 \pm 24$ mm & $625 \pm 13$ mm\\
	\hline
  Work  &  137 & $1256 \pm 14$ mm/s  &  $845 \pm 19$ mm & $731 \pm 12$ mm\\
	\hline
\end{tabular}
\end{center}
\end{table}
\subsubsection{The effect of relation}
\begin{figure}[tb!]
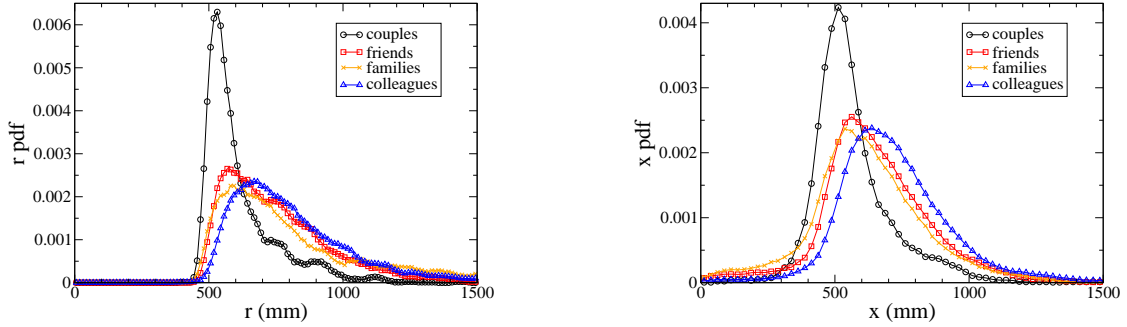

\begin{center}
\includegraphics[width=0.4\linewidth]{f12a.eps} \hspace{0.1\linewidth} \includegraphics[width=0.4\linewidth]{f12b.eps}
\caption{Probability distribution function for the $r$ (left) and $x$ (right) variables in groups depending on their relation.}
\label{f12}
\end{center}
\end{figure}

Table \ref{T3} compares the values of all observables between groups whose apparent relation was: couples, friends, family or colleagues; while Fig. \ref{f12} shows the probability distributions of the corresponding 
spatial variables.
\begin{table}[!ht]
\small
\caption{Observable dependence on relation for 2 people groups.}
\label{T3}
\begin{center}
\begin{tabular} {|c|c|c|c|c|}
	\hline
       &  $N$ &  $V$ &    $r$ & $x$  \\
	\hline
  Couples &  24 & $1068 \pm 41$ mm/s  &  $644 \pm 27$ mm & $597 \pm 27$ mm\\
	\hline
  Friends  &  56 & $1069 \pm 32$ mm/s  &  $785 \pm 26$ mm & $654 \pm 22$ mm\\
	\hline
  Families &  52 & $1055 \pm 24$ mm/s  &  $898 \pm 50$ mm & $609 \pm 20$ mm\\
	\hline
  Colleagues  &  140 & $1253 \pm 14$ mm/s  &  $842 \pm 18$ mm & $727 \pm 12$ mm\\
	\hline
\end{tabular}
\end{center}
\end{table}
\subsection{Discussion}
We may see that gender has a significant influence on group velocity and pedestrian distances, with females walking slower and closer than males. A similar statement can be done for purpose, with workers walking significantly faster and at a larger distance than people visiting the area for leisure. It may be nevertheless noticed that while the difference
in abreast distance $x$ corresponds to a few confidence values $\epsilon$, such a difference is more reduced for the absolute distance $r$. Regarding relation between pedestrians, we may see that couples, friends and family members walk at similar velocities, while colleagues walk much faster, suggesting that purpose affects velocity more than relation.
Relation has, on the other end, a strong effect on distance. Couples walk very close and abreast (the $x$ pdf assumes almost zero under 0.3 meters, suggesting that couples are almost always abreast), while friends walk at
a larger distance but much closer than colleagues. Families have peak distributions similar to friends, but have also a strong tendency not to walk abreast, due probably to the erratic behaviour of children.

There is clearly a strong overlapping between, for example, the populations of colleagues, workers and males. In order to better understand the contribution of these factors, in our upcoming work we will further 
break up these cathegories (``leisure oriented males'', etc.).
\section{Conclusion}
In this paper we exposed our mathematical theory of pedestrian group behaviour, provided new evidence regarding the non-Newtonian nature of pedestrian interactions, and performed a preliminary analysis, based on empirical observations,
 on the effect of group composition on dynamics. We observed that females walk slower and closer than males, that workers walk faster, at large distance and more abreast than leisure oriented people, and that
inter group relation has a strong effect on group structure, with couples walking very close and abreast, colleagues walking at a larger distance, and friends walking more abreast than family members.
\section{Acknowledgements}
This work was supported by CREST, JST and JSPS KAKENHI Grant Number 16J40223.


\begin{thebibliography}{99}
\bibitem{M}  F. Zanlungo, T. Ikeda, T. Kanda, {\it Potential for the dynamics of pedestrians in a socially interacting group}, Phys. Rev. E, 89, 1 , 021811, (2014) 
\bibitem{M2} F.~Zanlungo, D.~Br\v{s}\v{c}i\'c and T.~Kanda, {\it Spatial-size scaling of pedestrian groups under growing density conditions} Physical Review E 91 (6), 062810 (2015)
\bibitem{OOC} D. Br\v{s}\v{c}i\'c, F. Zanlungo, T. Kanda, {\it Density and velocity patterns during one year of pedestrian tracking}, Pedestrian and Evacuation Dynamics 2014
\bibitem{schultz2} M. Schultz, L. R{\"o}{\ss}ger, F. Hartmut and B. Schlag, {\it Group dynamic behavior and psychometric profiles as substantial driver for pedestrian dynamics},
in {\it Pedestrian and Evacuation Dynamics 2012}, U. Weidmann, U. Kirsh and M. Schreckenberg, Vol II, pp. 1097-1111 (2014)
\bibitem{mou2} M. Moussa{\"\i}d, N. Perozo, S. Garnier, D. Helbing and G. Theraulaz, {\it The walking behaviour of pedestrian social groups and its impact on crowd dynamics},
PLoS One, 5, 4, e10047, (2010)
\bibitem{costa} M. Costa, {\it Interpersonal distances in group walking}, Journal of Nonverbal Behavior, 34, 1, 15-26 (2010)
\bibitem{zan3} F. Zanlungo and T. Kanda, {\it Do walking pedestrians stabily interact inside a large group? Analysis of group and sub-group spatial structure},
COGSCI13, (2013)
\bibitem{dyads} A. Gorrini, G. Vizzari, S. Bandini, {\it Granulometric Distribution and Crowds of Groups: Focusing On Dyads}, 11th Conference International Traffic and Granular Flow, Delft (NL), 2015
\bibitem{koster} G. K{\"o}ster, M. Seitz, F. Treml, D. Hartmann and W. Klein, Contemporary Social Science, 6, 3, 397-414, (2011)
\bibitem{kara} I. Karamouzas, and M. Overmars, {\it Simulating the local behaviour of small pedestrian groups}, Proceedings of the 17th ACM Symposium on Virtual Reality Software and Technology,
183-190, (2010)
\bibitem{zhang} Y. Zhang, J. Pettr{\'e}, X. Qin, S. Donikian and Q. Peng, {\it A Local Behavior Model for Small Pedestrian Groups}, 
Computer-Aided Design and Computer Graphics (CAD/Graphics), 2011 12th International Conference on, 275-281 (2011)
\bibitem{M3} F.~Zanlungo and T.~Kanda, {\it A mesoscopic model for the effect of density on pedestrian group dynamics} Europhysics Letters, 111, 38007 (2015)
\bibitem{knapp} M. Knapp, {\it Nonverbal communication in human interaction} (Cengage Learning, Stamford, 2012)
\bibitem{kleinke} C. Kleinke, {\it Gaze and eye contact: a research review}, Psychological bulletin, 100, 1, 78 (1986)
\bibitem{Z} Z. Y{\"u}cel, F. Zanlungo, T. Ikeda, T. Miyashita, and N. Hagita {\it Deciphering the Crowd: Modeling and Identification of Pedestrian Group Motion}, Sensors, vol. 13, pp. 875-897, 2013
\bibitem{Turchetti} G. Turchetti, F. Zanlungo, B. Giorgini, {\it Dynamics and thermodynamics of a gas of automata}
EPL (Europhysics Letters) 78 (5), 58003
\bibitem{Hel} D. Helbing and P. Molnar, {\it Social force model for pedestrian dynamics}, Phys. Rev. E, 51, 5, 4282 (1995)
\bibitem{zan4} F. Zanlungo, T. Ikeda and T. Kanda, {\it Social force model with explicit collision prediction}, EPL, 93, 6, 68005 (2011)
\bibitem{dylan} D. Glas, T. Miyashita, H. Ishiguro and N. Hagita, {\it Laser-based tracking of human position and orientation using parametric shape modeling}, Advanced robotics, 23, 4, 405-428 (2009)
\bibitem{data} http://www.irc.atr.jp/sets/groups/
\bibitem{Drz} D. Br\v{s}\v{c}i\'c, T. Kanda, T. Ikeda, T. Miyashita {\it Person Tracking in large public spaces using 3-D range sensors}, IEEE Transactions on Human Machine Systems, vol 43, no 6, pp 522-534, (2013)

\end{thebibliography}
\end{document}